\documentclass[conference]{IEEEtran}
\usepackage[noadjust]{cite}
\usepackage{graphicx}
\usepackage[cmex10]{amsmath}
\usepackage{commath}
\usepackage{url}
\usepackage{graphicx}
\usepackage{amssymb}
\everymath{\displaystyle}

\usepackage[shortlabels]{enumitem}
\usepackage{nameref}
\usepackage{hyperref}
\usepackage{tikz}
\usepackage{esint}
\usepackage[shortcuts]{extdash}
\usepackage{comment}

\DeclareMathOperator{\sphericaldistance}{SphericalDistance}
\DeclareMathOperator{\sphericaldiskarea}{DiskArea}
\DeclareMathOperator{\sphericalsectorarea}{CircularSectorArea}

\DeclareMathOperator{\sphericalrighttrianglearea}{RightTriangleArea}
\DeclareMathOperator{\sphericalsegmentarea}{CircularSegmentArea}
\DeclareMathOperator{\intersectionarea}{IntersectionArea}

\newcommand{\kernelfunction}[2][]{{\operatorname{k}^{#1}}\left( #2 \right)}
\newcommand{\mycovariancefunction}[1]{{\operatorname{C}}\left( #1 \right)}
\newcommand{\expectation}[1]{{\operatorname{E}}\left[ #1 \right]}
\newcommand{\var}[1]{{\operatorname{Var}}\left[ #1 \right]}
\newcommand{\cov}[2]{{\operatorname{Cov}}\left[ #1 , #2 \right]}
\newcommand{\randomnormalizednoise}[1]{{\operatorname{W}}\left( #1 \right)}

\onecolumn

\begin{document}
\title
{
  New Flexible Compact Covariance Model\\
  on a Sphere
}
\author
{
  \IEEEauthorblockN{Alexander Gribov, Konstantin Krivoruchko}
  \IEEEauthorblockA
  {
    Esri\\
    380 New York Street\\
    Redlands, CA 92373\\
    Emails: agribov@esri.com; kkrivoruchko@esri.com
  }
}

\maketitle

\begin{abstract}
We discuss how the kernel convolution approach can be used to accurately approximate the spatial covariance model on a sphere using spherical distances between points. A detailed derivation of the required formulas is provided. The proposed covariance model approximation can be used for non-stationary spatial prediction and simulation in the case when the dataset is large and the covariance model can be estimated separately in the data subsets.
\end{abstract}




\section{Introduction}

To perform kriging on a sphere, one can use ``great arc/circle'' distance metrics (the shortest distance on the surface of a sphere) to compute distances between points and the covariance model which is positive definite on a sphere. Among the valid covariance models are spherical, stable with shape parameter $\alpha \leq 1$, and K-Bessel (same as Matern) with shape parameter $\alpha \leq \dfrac{1}{2}$, see, for example, \cite{Reference5KBesselOnSphere}. Disadvantages of this approach include the following:
\begin{itemize}[\textbullet]
  \item The allowed covariance models are not flexible because they change rapidly at short distances.
  \item These covariance models are isotropic. However, the analysis of various data has shown that an assumption of isotropy is generally inappropriate for global data because most of the geographical data are not stationary in latitude \cite{Reference6AnisotropicProcessOnSphere}.
  \item It is not necessarily true that the best model on the plane is also the best on the surface of a sphere and it seems important to derive new flexible covariance models which are valid on a sphere and correspond to particular physical processes.
\end{itemize}

A covariance model with zero values when the distance between two points exceeds specified threshold value has a compact support. Compactly supported covariance functions allow for computationally efficient sparse matrix techniques usage, which is especially important when dealing with large datasets. \cite{CompactCovariance} provides comprehensive review on the compact covariance construction for both stationary and non-stationary models on the plane and develops non-stationary compactly supported covariance functions for the spherical family. One advantage of the spherical model is its closed-form expression. However, the model proposed in \cite{CompactCovariance} is not flexible enough because it has a fixed shape. \cite{KernelConvolutionForRingsOnPlane} shows how flexible compact covariance model can be approximated numerically for further efficient use in prediction and simulation. In this paper we propose flexible compact covariance model which describes both weak and strong spatial correlation on a sphere. If required, non-stationary model can be constructed as shown in the conclusion section below.

Some of the issues listed above can be resolved by using the ker\-nel\-/con\-vo\-lu\-tion approach. Taking into account that the shortest distance on the unit sphere is the angle (in radians) between two locations on a plane passing through the sphere center and the two points, we define the following kernel function:
\begin{equation}
  \kernelfunction{ h | \mu, \nu }
  =
  \left\{
    \begin{aligned}
      & \left( 1 - h ^ \mu \right) ^ \nu, \text{if } h < 1,\\
      & 0, \text{otherwise},
    \end{aligned}
  \right.
  \label{eq:SphericalKernelWithTwoParameters}
\end{equation}
where $ \mu > 0 $ and $ \nu > 0 $ are the shape parameters and $h \geq 0$ is the distance from the center of the kernel.\footnote{Vertical bar symbol $|$ separates variables from parameters.}

On a sphere, kernel $\kernelfunction{ h | \mu, \nu }$ can be represented as
\begin{equation*}
  \kernelfunction{ s | s_*, r, \mu, \nu }
  =
  \dfrac
  {
    \kernelfunction{ \left. \frac{2 \cdot \sphericaldistance \left( s, s_* \right)}{r} \right| \mu, \nu }
  }
  {
    \sqrt
    {
      \oiint
      {
        k^2{\left( \left. \frac{2 \cdot \sphericaldistance \left( s', s_* \right)}{r} \right| \mu, \nu \right)}
        \cdot
        \dif s'
      }
    }
  }
  ,
\end{equation*}
where
$s_*$ is the location of the kernel center on a sphere,
$s$ is the location where kernel value is calculated,
$\dfrac{r}{2}$ is radius of the kernel (in radians),
$\sphericaldistance$ is the shortest distance on the surface of a sphere, see formula \eqref{EquationSphericalDistance} in section~\ref{sec:SectionMathematicalFormulas} below. The integration ($\oiint$) is performed on the surface of a sphere.

Kernel \eqref{eq:SphericalKernelWithTwoParameters} generates the following random process on a sphere:
\begin{equation*}
  Y \left( s \right)
  =
  \oiint
  {
    \kernelfunction{ s' | s, r, \mu, \nu }
    \cdot
    \dif \randomnormalizednoise{s'}
  }
  ,
\end{equation*}
where
$ \randomnormalizednoise{s'} $ is a diffusion process in three dimensions,
$ \dif \randomnormalizednoise{s'} $ is white noise,
$
  \expectation
  {
    \oiint
    \limits
    _A
    {
      \dif \randomnormalizednoise{s'}
    }
  }
  =
  0
$,
and
$
  \var
  {
    \oiint
    \limits
    _A
    {
      \dif \randomnormalizednoise{s'}
    }
  }
  =
$
$
  \oiint
  \limits
  _A
  {
    \dif s'
  }
$
$
  \forall
  A
$,
where
$
  A
$
is some integrable area on a sphere.

The relationship between the kernel and the covariance function is the following:
\begin{equation*}
  \cov{Y \left( s_1 \right)}{Y \left( s_2 \right)}
  =
  \oiint
  {
    \kernelfunction{ s' | s_1, r, \mu, \nu }
    \cdot
    \kernelfunction{ s' | s_2, r, \mu, \nu }
    \cdot
    \dif s'
  }
  .
\end{equation*}

Note that a kernel with radius $\dfrac{r}{2}$ generates the covariance function with a range equal to $r$ and the covariance depends only on the angle $\alpha$ between points,
\begin{equation}
  \mycovariancefunction{\alpha | r, \mu, \nu}
  =
  \oiint
  {
    \kernelfunction{ s' | (1, 0, 0), r, \mu, \nu }
    \cdot
    \kernelfunction{ s' | (\cos \left( \alpha \right), \sin \left( \alpha \right), 0), r, \mu, \nu }
    \cdot
    \dif s'
  }
  .
  \label{eq:SphericalCovarianceFunction}
\end{equation}

The covariance shape is a function of the range and parameters $\mu$ and $\nu$ in contrast to the covariance defined on the plane, which is a function of the parameters $\mu$ and $\nu$ only. This feature of the covariance on a sphere is important for the calculations optimization.

Unfortunately, not all kernel functions are integrable. Moreover, numerical evaluation of the positive-definite property of the covariance is problematic because it should be done with high precision. In the case of approximate integration, there is no guarantee that the resulting covariance model is positive definite and that the model describes the same stochastic process as the model which is calculated exactly.

One recent attempt to find an approximate solution of the problem was made in \cite{Refernce13CovarianceIntegration}. The author of \cite{Refernce13CovarianceIntegration} assumes that the parameters of the random field at each latitude are homogeneous and the modeling process at each latitude is isotropic. The kernel was chosen in such a way that its convolution gives covariance with the shape similar to the Matern one (however, for simplicity, it was assumed that the shape parameter is fixed and constant on the entire globe). It was also assumed that the data are absolutely precise so that the nugget effect is zero. Another assumption was that the sill and range are constants at each latitude, but they are changing smoothly between the latitudes. Then the values of the sill and range were found in one dimension for a fixed number of latitudes and a linear smoother was used to interpolate these estimates. \mbox{Finally}, the convolution of kernels was found numerically using the sets of discrete values inside the predefined ranges of each kernel parameter and it was reported that the approximation error was less than $0.01$.

\begin{figure} [htb]
	\centering
	\includegraphics[width = 0.6\columnwidth, keepaspectratio]{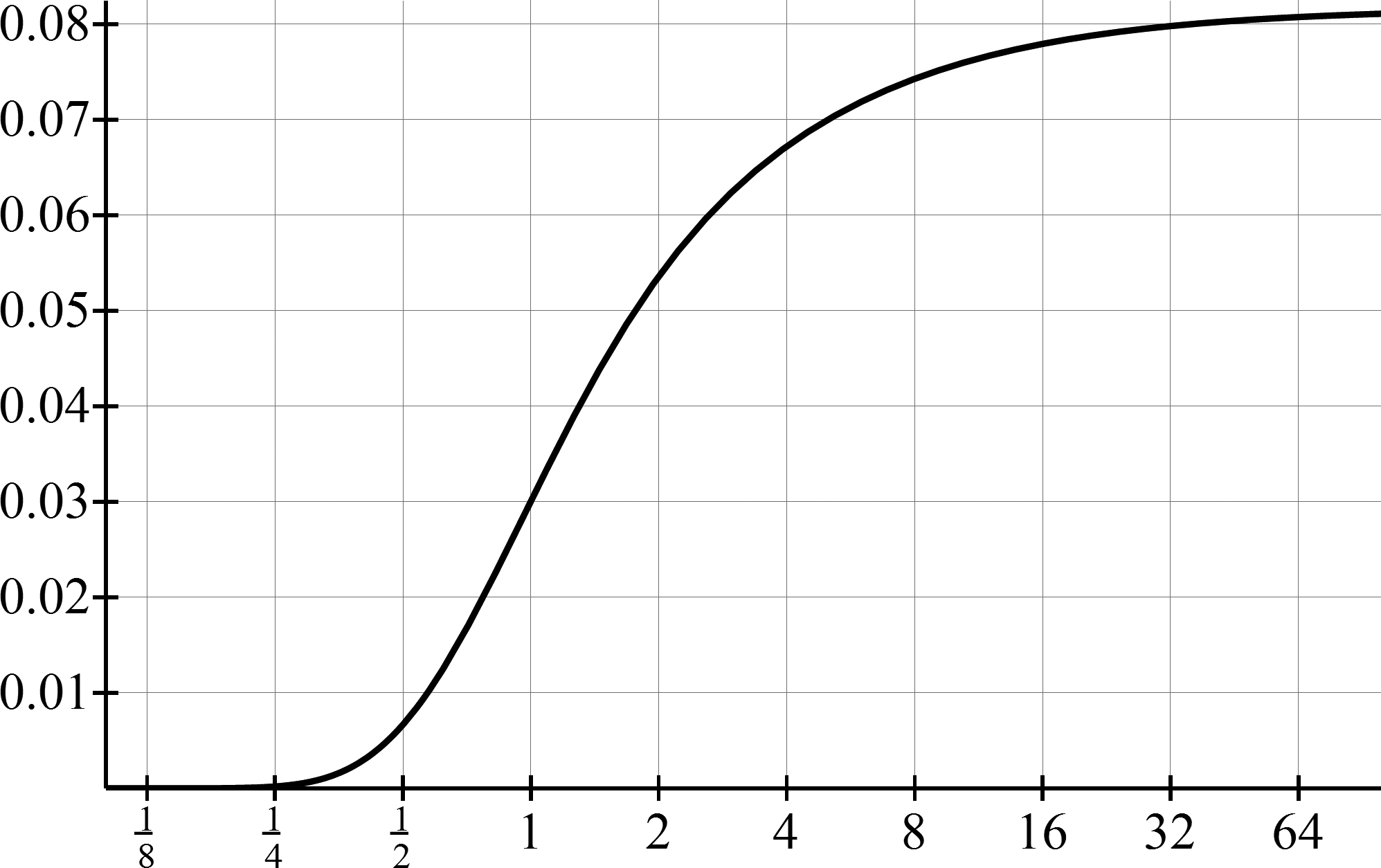}
	\caption{The proportion of effective range at which calculation of the covariance is not possible (axis~Y) for the shape parameter of the Mattern covariance (axis~X).}
	\label{fig:DeterminantZeroGraph}
\end{figure}

However, the error of $0.01$ is too large for calculation of the covariance model. Consider the Matern covariance with effective range of $1$. The graph in Figure~\ref{fig:DeterminantZeroGraph} shows the proportion of effective range at which calculation of the covariance is not possible due to the inaccuracy in the covariance matrix (when the determinant equals zero) as a function of the shape parameter. The error of $0.01$ will lead to serious problem when the samples are separated by relatively small distances. For example, when the effective range is $100$~km, the covariance values cannot be calculated for the distances between points less than $\approx\!\!6$~km for typical values of the shape parameter (in fact, these critical distances will be even larger because of numerical instability near the threshold at which the determinant equals zero). Therefore, the proposed approximation in \cite{Refernce13CovarianceIntegration} can be used only in a very special case when the error-free data are nearly regularly sampled and the effective range of the data correlation is $10$~- $20$ times larger than the distance between the neighboring samples.

A model which overcomes most of the above-men\-tioned problems can be constructed based on the ker\-nel-con\-vo\-lu\-tion approach proposed in \cite{Reference1bKernelConvolution} and \cite{Reference1aKernelConvolution}. In particular, \cite{Reference1bKernelConvolution} showed that the integral can be calculated analytically for any kernel step function on the plane. That methodology is also valid in larger dimensions.

\begin{figure*} [p]
	\centering
	\begin{tabular}{c}
		\includegraphics[width = 16.5 cm, keepaspectratio]{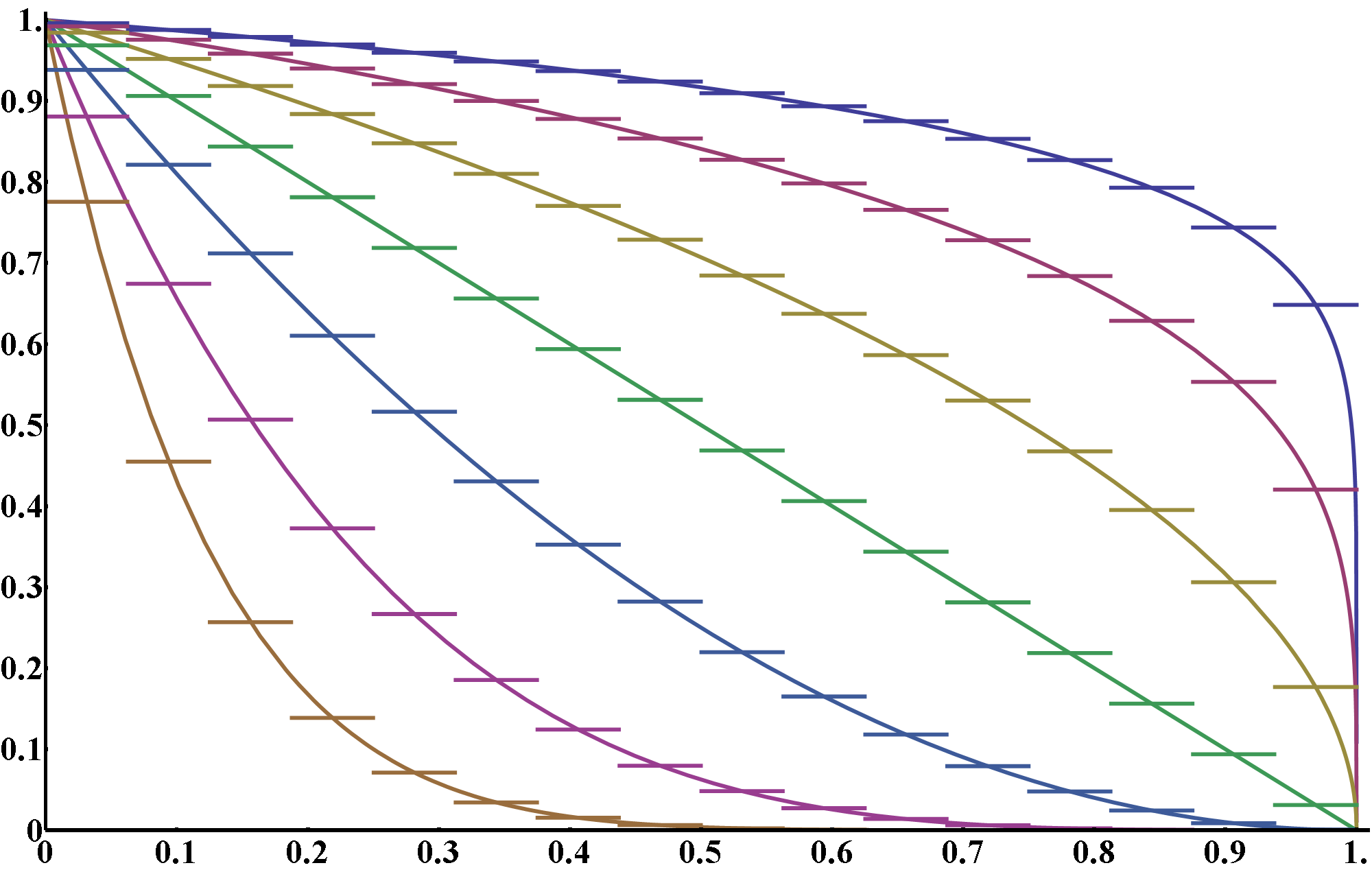}
		\\
		a)
		\\
		\includegraphics[width = 16.5 cm, keepaspectratio]{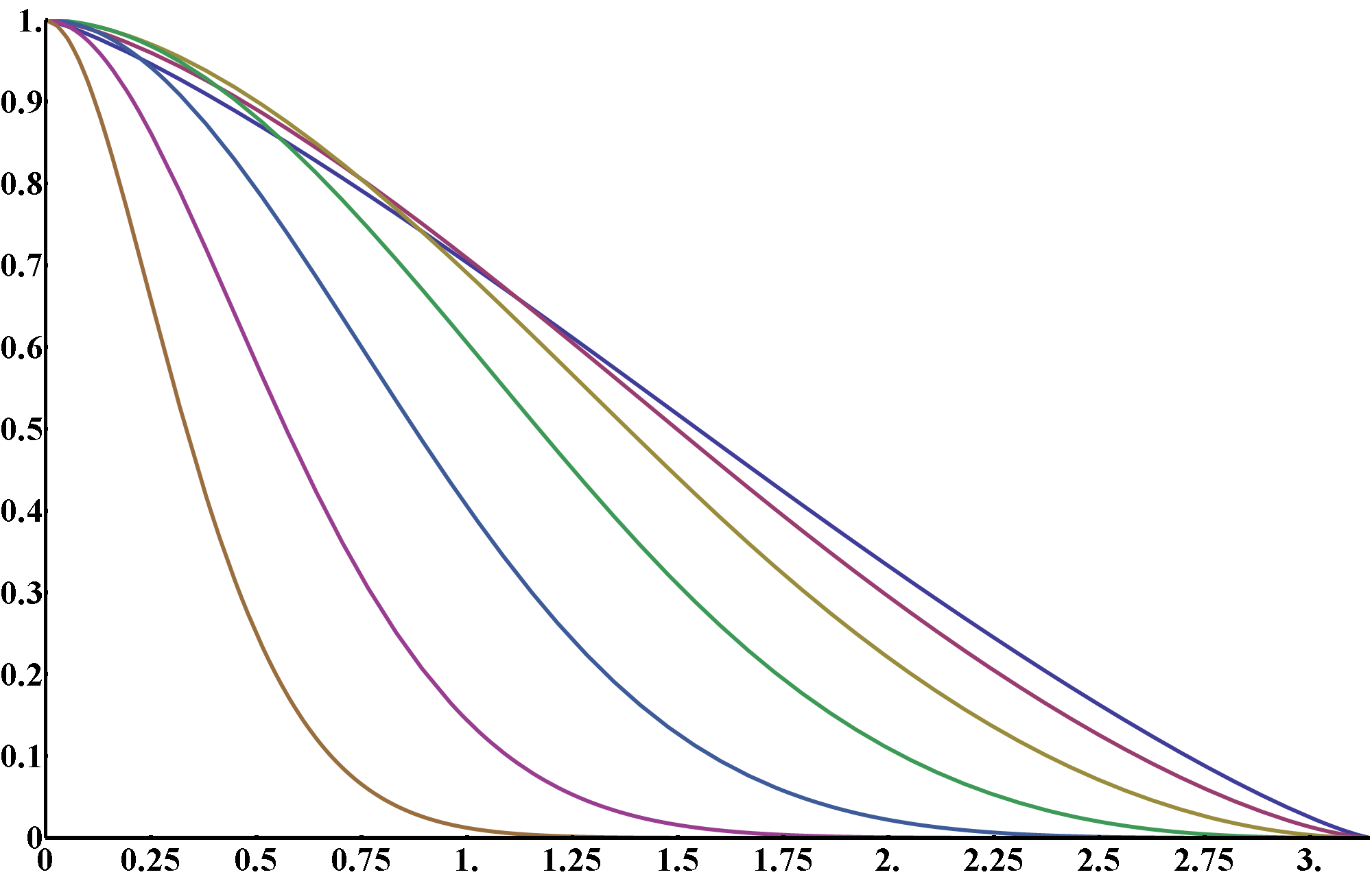}
		\\
		b)
	\end{tabular}
	\caption
	{
		Approximation of smooth kernel \eqref{eq:SphericalKernelWithTwoParameters} by step function with $16$ steps (a) and covariance function with range $\pi$ constructed by kernels with $64$ steps (b) for $\mu = 1$ and $\nu = \dfrac{1}{8}$, $\dfrac{1}{4}$, $\dfrac{1}{2}$, $1$, $2$, $4$, and $8$.
	}
	\label{fig:ApproximationKenrenByStepFunctionAndCovarianceFunctionWithLargeNumberOfSteps}
\end{figure*}

Figure~\ref{fig:ApproximationKenrenByStepFunctionAndCovarianceFunctionWithLargeNumberOfSteps}a shows the step functions for several kernels $\kernelfunction{ \cdot | \mu, \nu }$ defined in \eqref{eq:SphericalKernelWithTwoParameters}. The larger the number of steps, the more accurate the approximation of the kernel. Figure~\ref{fig:ApproximationKenrenByStepFunctionAndCovarianceFunctionWithLargeNumberOfSteps}b shows a set of calculated flexible covariance models with the range of data correlation equal to $\pi$, the maximum distance between points on a sphere.

Numerically calculated covariance can be fitted using one of the available algorithms. We use restricted maximum likelihood, but other fitting algorithms, including weighted least squares, can be used as well.


In the next section, we show how the covariance can be produced by the kernel convolution with a kernel step function on a sphere. Formulas in the next section can be generalized to allow change of the kernel size and shape.

In the conclusion section, we discuss prediction and simulation with the proposed compactly supported covariance model on a sphere for both stationary and non-stationary data.

\section
{
  Derivation of the covariance on a sphere produced by kernel convolution
  \label{sec:SectionMathematicalFormulas}
}

In this section, we provide definitions and derivations required for calculation of the covariance on a unit sphere located at the origin of the coordinate system. Detailed information on the spherical geometry can be found, for example, in \cite{SphericalTrigonometry}. Note that all angles in the formulas below are in radians.

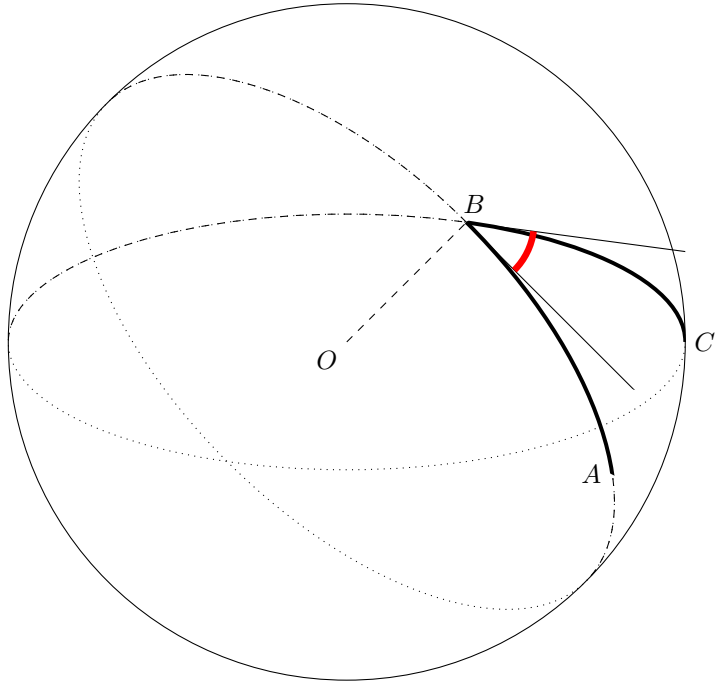
\begin{figure} [htb]
  \centering
  \begin{tikzpicture}[scale = 4.5] 
    \def\MySphericalAngleRadius{0.2}
  
    \draw (0, 0) circle (1);
  
  
  
    \draw [dashed] (0, 0) -- (0.35355339059327376220042218105242, 0.35355339059327376220042218105242);
  
    \draw (0, 0) node [anchor = north east] {$O$};
    \draw (0.78, -0.39) node [anchor = east] {$A$};
    \draw (0.35355339059327376220042218105242, 0.35355339059327376220042218105242) node [anchor = 247.5] {$B$};
    \draw (1, 0) node [anchor = west] {$C$};
  
    \draw [dotted, rotate = 45] (0, 0) ellipse (1 / 2 and 1);
    \begin{scope}[even odd rule]
      \clip (1, -1) -- (-1, 1) -- (1, 1) -- cycle;
      \draw [dashed, rotate = 45] (0, 0) ellipse (1 / 2 and 1);
    \end{scope}
    \begin{scope}[even odd rule]
      \clip (0, 0) -- (1, -1 / 2) -- (1, 1) -- cycle;
      \draw [line width = 1.5, rotate = 45] (0, 0) ellipse (1 / 2 and 1);
    \end{scope}
    \begin{scope}[even odd rule]
      \clip (0, 0) -- (1, -1 / 6) -- (1, 1) -- cycle;
      \draw [rotate = 45] (1 / 2, 0) -- (1 / 2, -1);
    \end{scope}
  
    \draw [dotted] (0, 0) ellipse (1 and 0.37796447300922722721451653623418);
    \begin{scope}[even odd rule]
      \clip (-1, 0) -- (-1, 1) -- (1, 1) -- (1, 0) -- cycle;
      \draw [dashed] (0, 0) ellipse (1 and 0.37796447300922722721451653623418);
    \end{scope}
    \begin{scope}[even odd rule]
      \clip (0, 0) -- (1, 1) -- (1, 0) -- cycle;
      \draw [line width = 1.5] (0, 0) ellipse (1 and 0.37796447300922722721451653623418);
    \end{scope}
    \begin{scope}[even odd rule]
      \clip (0, 0) -- (1, 1) -- (1, 0) -- cycle;
      \draw (0.35355339059327376220042218105242, 0.35355339059327376220042218105242) -- (0.35355339059327376220042218105242 + 2.6457513110645905905016157536393, 0.35355339059327376220042218105242 - 0.35355339059327376220042218105242);
    \end{scope}
  
    \draw [line width = 2.5, red] (0.35355339059327376220042218105242 + 0.99118925556670416976085082980775 * \MySphericalAngleRadius, 0.35355339059327376220042218105242 - 0.13245323570650438065327517205169 * \MySphericalAngleRadius) arc (-7.611378517094656748635568675606: -45: \MySphericalAngleRadius);
  \end{tikzpicture}
  \caption[]
  {
    Spherical angle $\sphericalangle{ABC}$, shown as thick red arc, between two spherical segments $AB$ and $BC$.
  }
  \label{fig:SphericalAngleSphericalRightTriangleAndAreaOfSphericalCircularSegmentA}
\end{figure}

\begin{itemize}[\textbullet]
  \item A spherical segment is the shortest path connecting two points on the surface of a sphere. It coincides with the great arc passing through those two points.

  \item \label{DefinitionDisanceOnTheSphere} A spherical distance between points $\left( x_1, y_1, z_1 \right)$ and $\left( x_2, y_2, z_2 \right)$ on a sphere is
  \begin{multline}
  \sphericaldistance{\left( \left( x_1, y_1, z_1 \right) , \left( x_2, y_2, z_2 \right) \right)}
  =
  \arccos
  {
    \left( x_1 \cdot x_2 + y_1 \cdot y_2 + z_1 \cdot z_2 \right)
  }
  =
  \\
  =
  2
  \arcsin
  {
    \left(
      \tfrac
      {
        \sqrt
        {
          \left(
            x_2 - x_1
          \right)
          ^
          2
          +
          \left(
            y_2 - y_1
          \right)
          ^
          2
          +
          \left(
            z_2 - z_1
          \right)
          ^
          2
        }
      }
      {
        2
      }
    \right)
  }
  .
    \label{EquationSphericalDistance}
  \end{multline}

  This distance equals the angle between two points on a sphere from its center. The formula based on $\arcsin$ has higher precision for small angles.\footnote{Scalar product under $ \arccos $ has error proportional to the maximum absolute value of the coordinates while the distance under $ \arcsin $ has error proportional to the maximum absolute difference between the coordinates. As angle approaches zero, the absolute difference between the coordinates tends to zero.}

  \item A spherical triangle consists of spherical segments $AB$, $BC$, and $AC$, where $A$, $B$, and $C$ are points on a sphere.

  \item A spherical angle is an angle on the surface of a sphere between two spherical segments, see Figure~\ref{fig:SphericalAngleSphericalRightTriangleAndAreaOfSphericalCircularSegmentA}. To distinguish spherical angles from plane angles, the symbol $\sphericalangle$ is used.

  \item A spherical right triangle is a spherical triangle where one spherical angle equals $\dfrac{\pi}{2}$ (right angle).
\end{itemize}

Using the definitions above, several spherical geometry relationships between spherical segments and angles follows.

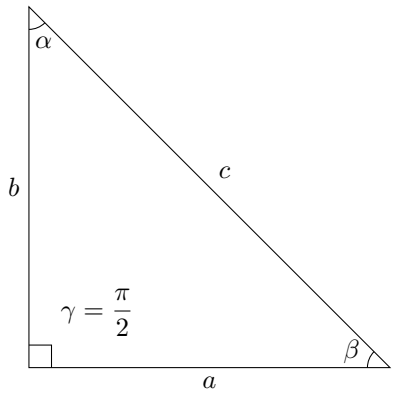
\begin{figure} [htb]
\centering
\begin{tikzpicture} [scale = 0.6] 
\def\MyFigureProportion{2}
\draw (0, 0) -- (8, 0) -- (0, 8) -- cycle;
\draw (4, 0) node [anchor = north] {$a$};
\draw (0, 4) node [anchor = east] {$b$};
\draw (4, 4) node [anchor = south west] {$c$};
\draw (1 / \MyFigureProportion, 0) -- (1 / \MyFigureProportion, 1 / \MyFigureProportion) -- (0, 1 / \MyFigureProportion);
\draw (1 / \MyFigureProportion, 1 / \MyFigureProportion) node [anchor = south west] {$\gamma = \dfrac{\pi}{2}$};
\draw (8 - 1 / \MyFigureProportion, 0) arc (180:135:1 / \MyFigureProportion);
\draw (8 - 0.9238795325112867561281831893968 / \MyFigureProportion, 0.3826834323650897717284599840304 / \MyFigureProportion) node [anchor = 337.5] {$\beta$};
\draw (0 / \MyFigureProportion, 8 - 1 / \MyFigureProportion) arc (270:315:1 / \MyFigureProportion);
\draw (0.3826834323650897717284599840304 / \MyFigureProportion, 8 - 0.9238795325112867561281831893968 / \MyFigureProportion) node [anchor = 112.5] {$\alpha$};
\end{tikzpicture}
\caption[]
{
	Spherical right triangle.
}
\label{fig:SphericalAngleSphericalRightTriangleAndAreaOfSphericalCircularSegmentB}
\end{figure}

\begin{enumerate}[label = (\emph{\Roman*}), align = left]
  \item \label{DefinitionPythagoreanTheorem} The relationship between sides $a$, $b$, and hypotenuse $c$ of spherical right triangle\footnote{$a$, $b$, and $c$ are spherical segments, see definitions for spherical segment and spherical distance, between two points, in the beginning of this section.} \cite{SphericalTrigonometry}, see Figure~\ref{fig:SphericalAngleSphericalRightTriangleAndAreaOfSphericalCircularSegmentB}, is:
  \begin{equation}
    \cos{\left( c \right)} = \cos{\left( a \right)} \cdot \cos{\left( b \right)}.
    \label{PythagoreanTheoremOnSphereReference}
  \end{equation}

  This is equivalent to
  \begin{equation}
    \sin^2{\left( \dfrac{c}{2} \right)}
    =
    \sin^2{\left( \dfrac{a}{2} \right)}
    +
    \sin^2{\left( \dfrac{b}{2} \right)}
    -
    2
    \sin^2{\left( \dfrac{a}{2} \right)}
    \cdot
    \sin^2{\left( \dfrac{b}{2} \right)}
    \label{PythagoreanTheoremOnSphere}
  \end{equation}

  \eqref{PythagoreanTheoremOnSphere} has better precision than \eqref{PythagoreanTheoremOnSphereReference} for small angles.

  This is equivalent to the Pythagorean theorem. Notice that \eqref{PythagoreanTheoremOnSphere} tends toward $c^2 = a^2 + b^2$, when $a$, $b$, and $c$ approach $0$.

  \item \label{DefinitionSphericalAngleOfSphericalRightTriangle} The relationship between spherical angle $\beta$, side $a$, and hypotenuse $c$ of spherical right triangle \cite{SphericalTrigonometry}, see Figure~\ref{fig:SphericalAngleSphericalRightTriangleAndAreaOfSphericalCircularSegmentB}, is:
  \begin{equation}
    \cos{\left( \beta \right)} = \tan{\left( a \right)} \cdot \cot{\left( c \right)}.
    \label{CosineTheoremOnSphere}
  \end{equation}

  This is equivalent to the definition of cosine for right triangles. Notice that \eqref{CosineTheoremOnSphere} tends toward $\cos \beta = \dfrac{a}{c}$, when $a$ and $c$ approach $0$.

\end{enumerate}

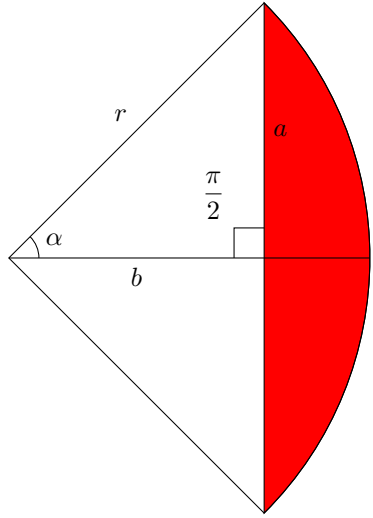
\begin{figure} [htb]
  \centering
  \begin{tikzpicture} [scale = 0.8] 
    \def\MyFigureProportion{2}
    \fill [fill = red] (6 * 0.70710678118654752440084436210485, -6 * 0.70710678118654752440084436210485) arc (-45:45:6) -- cycle;
    \draw (6 * 0.70710678118654752440084436210485, -6 * 0.70710678118654752440084436210485) arc (-45:45:6) -- cycle;
    \draw (0, 0) -- (6 * 0.70710678118654752440084436210485, -6 * 0.70710678118654752440084436210485) arc (-45:45:6) -- cycle;
    \draw (0, 0) -- (6, 0);
    \draw (1 / \MyFigureProportion, 0) arc (0:45:1 / \MyFigureProportion);
    \draw (0.9238795325112867561281831893968 / \MyFigureProportion, 0.3826834323650897717284599840304 / \MyFigureProportion) node [anchor = 202.5] {$\alpha$};
    \draw (6 * 0.70710678118654752440084436210485 - 1 / \MyFigureProportion, 0) -- (6 * 0.70710678118654752440084436210485 - 1 / \MyFigureProportion, 1 / \MyFigureProportion) -- (6 * 0.70710678118654752440084436210485, 1 / \MyFigureProportion);
    \draw (6 * 0.70710678118654752440084436210485 - 1 / \MyFigureProportion, 1 / \MyFigureProportion) node [anchor = south east] {$\dfrac{\pi}{2}$};
    \draw (3 * 0.70710678118654752440084436210485, 3 * 0.70710678118654752440084436210485) node [anchor = south east] {$r$};
    \draw (3 * 0.70710678118654752440084436210485, 0) node [anchor = north] {$b$};
    \draw (6 * 0.70710678118654752440084436210485, 6 * 0.70710678118654752440084436210485 / 2) node [anchor = west] {$a$};
  \end{tikzpicture}
  \caption[]
  {
    Spherical circular sector (see definition of planar circular sector in \cite{CircularSector}) with spherical circular segment (see definition of planar circular segment in \cite{CircularSegment}) area in red.
  }
  \label{fig:SphericalAngleSphericalRightTriangleAndAreaOfSphericalCircularSegmentC}
\end{figure}

\begin{enumerate}[label = (\emph{\Roman*}), resume, align = left]
  \item \label{DefinitionAreaOfSphericalTriangle} 
    %
    %
  The area of a spherical right triangle with sides $a$ and $b$, see Figure~\ref{fig:SphericalAngleSphericalRightTriangleAndAreaOfSphericalCircularSegmentB} and \cite{SphericalTrigonometry}, is:
  \begin{equation}
    \sphericalrighttrianglearea{\left( a, b \right)}
    =
    2
    \arctan
    {
      \left(
        \tan
        {
          \left(
            \dfrac{a}{2}
          \right)
        }
        \cdot
        \tan
        {
          \left(
            \dfrac{b}{2}
          \right)
        }
      \right)
    }
    .
    \label{AreaOfSphericalRightTriangle}
  \end{equation}

  Notice that \eqref{AreaOfSphericalRightTriangle} tends toward area of the right triangle $\dfrac{a \cdot b}{2}$, when $a$ and $b$ approach $0$.

  \item \label{DefinitionSphericalDisk} A spherical circle with radius $r$ at point $\left( x_c, y_c, z_c \right)$ is a set of points on the surface of a sphere with a spherical distance equal to $r$ from its center.

  A spherical disk with radius $r$ at point $\left( x_c, y_c, z_c \right)$ contains points on the surface of a sphere with a spherical distance less than or equal to $r$ from its center.

  A spherical ring is the area of the outer spherical disk excluding the inner spherical disk sharing the same center.

  \item \label{DefinitionMaximumSphericalDiskRadius} It follows that a maximum radius of a spherical disk is $\pi$. That spherical disk covers a sphere completely.

  \item \label{DefinitionComplementOfAnySphericalDisk} We ignore the edge of the spherical disk because it does not affect its area. The complement of a spherical disk with radius $r$ and center $\left( x_c, y_c, z_c \right)$ is a spherical disk with center at the opposite point on the surface of a sphere $\left( -x_c, -y_c, -z_c \right)$ with radius equal to $\pi - r$.

  \item \label{DefinitionComplementOfAnySphericalDiskMinimumRadius} From \ref{DefinitionMaximumSphericalDiskRadius} and \ref{DefinitionComplementOfAnySphericalDisk}, it follows that a minimum radius of a spherical disk and its complement is less than or equal~to~$\dfrac{\pi}{2}$.

  \item \label{DefinitionAreaOfSphericalDisk} The area inside the spherical disk with radius $r$ is
  \begin{equation}
    \sphericaldiskarea{\left( r \right)}
    =
    2 \pi \cdot (1 - \cos{\left( r \right)})
    =
    4 \pi \cdot \sin^2{\left( \dfrac{r}{2} \right)}.
    \label{SphericalCircleArea}
  \end{equation}
  
  The area of the spherical circular sector, defined in Figure~\ref{fig:SphericalAngleSphericalRightTriangleAndAreaOfSphericalCircularSegmentC}, with sector spherical angle $\alpha$ and radius $r$ is
  \begin{equation*}
    \sphericalsectorarea{\left( \alpha, r \right)}
    =
    \alpha \cdot (1 - \cos{\left( r \right)})
    =
    2 \alpha \cdot \sin^2{\left( \dfrac{r}{2} \right)}.
  \end{equation*}
\end{enumerate}

\begin{enumerate}[label = (\emph{\Roman*}), resume, align = left]
  \item \label{DefinitionSphereArea} From \ref{DefinitionMaximumSphericalDiskRadius} and \eqref{SphericalCircleArea}, the area of a sphere is $4 \pi$.

  \item \label{DefinitionAreaOfSphericalCircularSegment} The area of the spherical circular segment, see Figure~\ref{fig:SphericalAngleSphericalRightTriangleAndAreaOfSphericalCircularSegmentC}, is
  \begin{equation*}
    \sphericalsegmentarea{\left( \alpha, r, a, b \right)}
    =
    2
    \left(
      \sphericalsectorarea{\left( \alpha, r \right)}
      -
      \sphericalrighttrianglearea{\left( a, b \right)}
    \right)
    .
  \end{equation*}
\end{enumerate}

  Now we have all necessary formulas for the covariance calculations. First, we define the kernel step function as
  \begin{equation}
    \kernelfunction{ h }
    =
    \sum\limits_{j = 1}^{n}{\left( a_j \cdot \delta_{r_{j - 1} \leq h \wedge h < r_{j}} \right)}
    =
    \sum\limits_{j = 1}^{n}{\left( b_j \cdot \delta_{h < r_{j}} \right)}
    ,
    \label{KernelFunctionUnNormalized}
  \end{equation}
  where
  $n$ is the number of steps in the kernel function,
  $a_j$ is the value of the kernel function at step $ j $, $j = \overline{1..n}$,
  $b_j = a_j - a_{j + 1}$, $j = \overline{1..n - 1}$ and $b_n = a_n$,
  $r_{j}$ is a sequence of steps, $j = \overline{0..n}$ ($0 = r_0 < r_1 < r_2 < ... < r_{n} \leq \pi$),
  $\delta_{\theta}$ is $1$ when $\theta$ is true and $0$ otherwise.

  The kernel step function located at point $\left( x_c, y_c, z_c \right)$ is
  \begin{equation*}
    \kernelfunction{ \left( x, y, z \right) | \left( x_c, y_c, z_c \right) }
    =
    \kernelfunction{ \sphericaldistance \left( \left( x, y, z \right) , \left( x_c, y_c, z_c \right) \right) }
    .
  \end{equation*}

\begin{figure*} [htb]
  \centering
  \begin{tabular}{c c c c}
    \includegraphics[scale = 0.3]{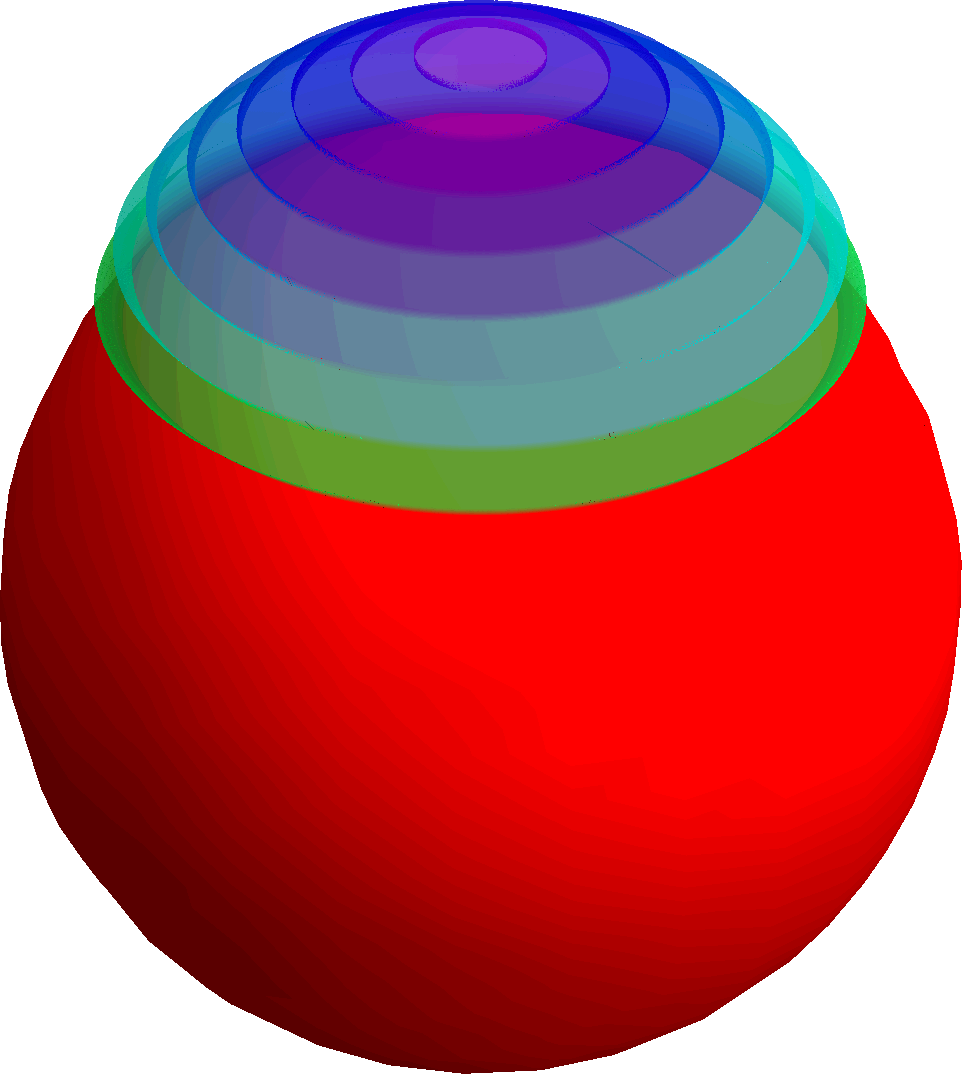}
    &
    &
    &
    \includegraphics[scale = 0.3]{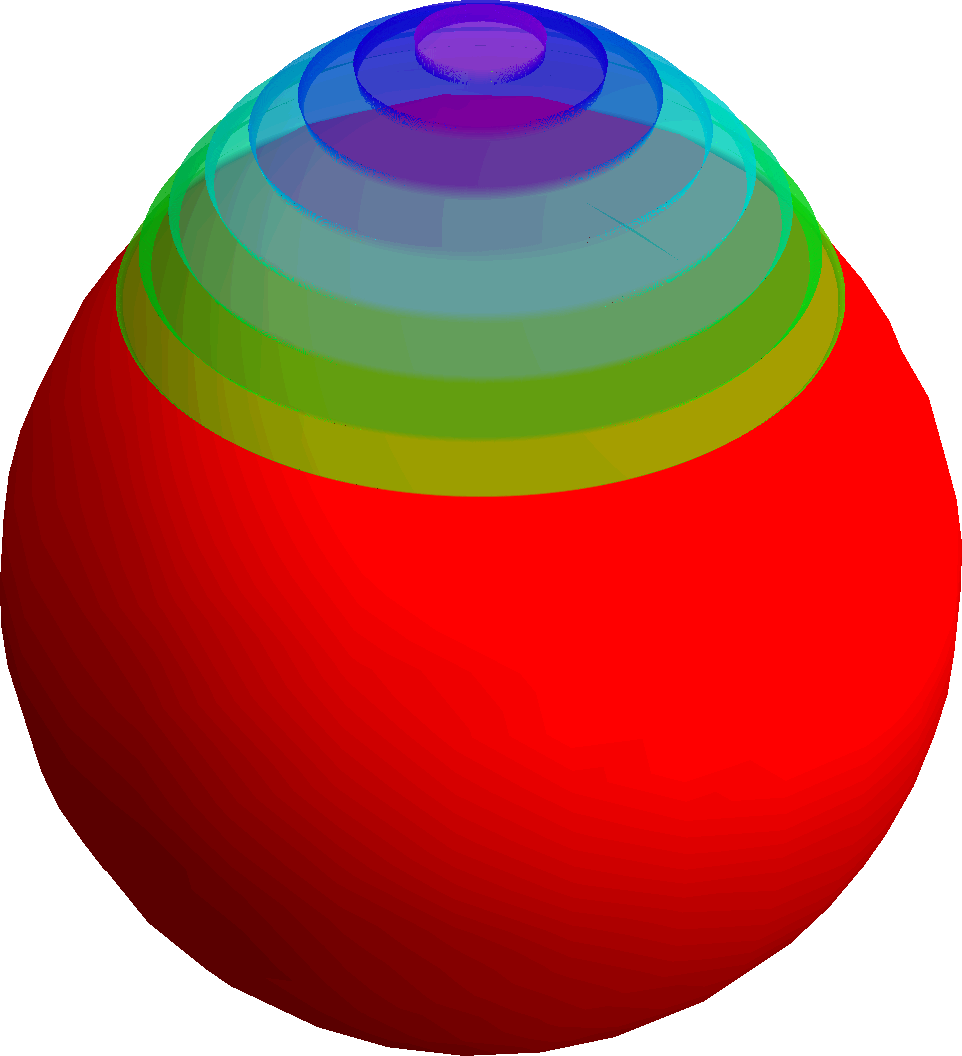}
    \\
    \\
    \\
    \includegraphics[scale = 0.3]{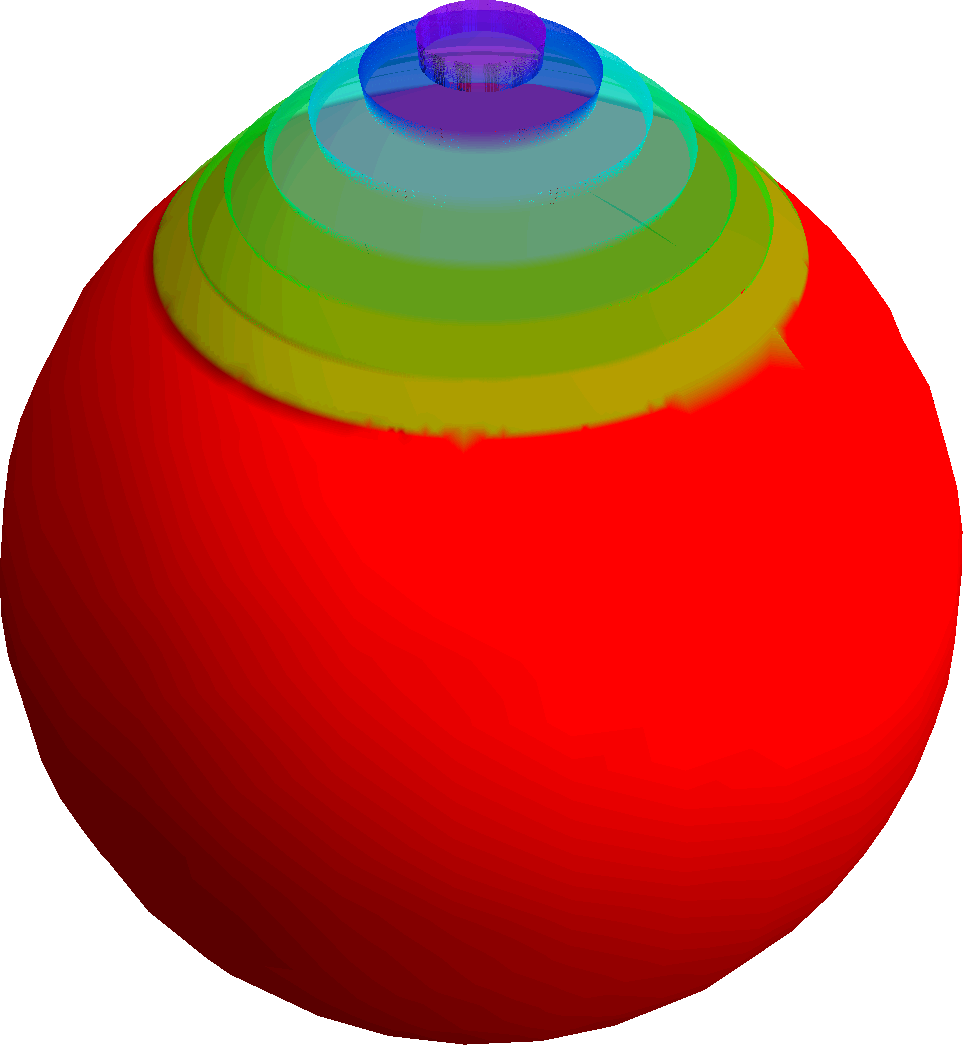}
    &
    &
    &
    \includegraphics[scale = 0.3]{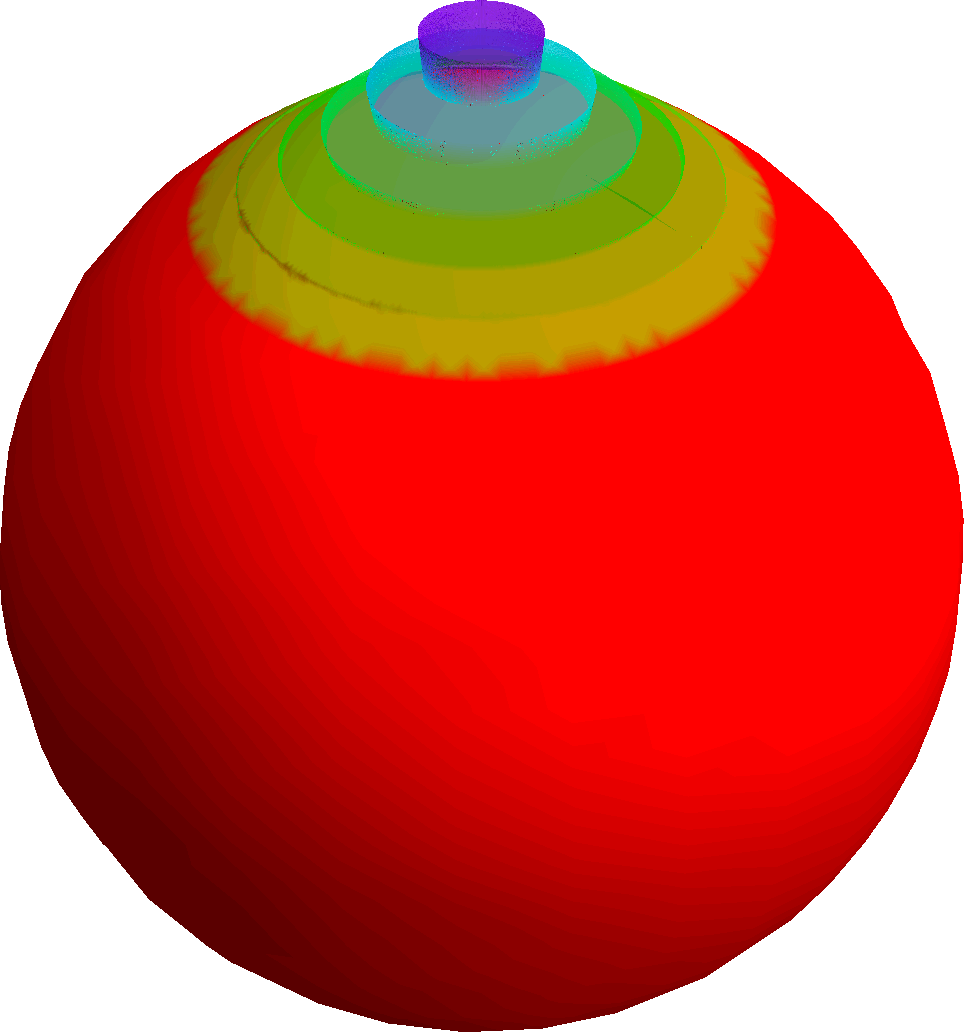}
    \\
  \end{tabular}
  \caption{Four kernel step functions located at the top of a sphere.}
  \label{fig:StepKernelExample}
  \end{figure*}

  Figure~\ref{fig:StepKernelExample} shows four different kernels on a sphere.

  To construct the covariance function with variance equal to $1$, the kernel function at any location $\left( x_c, y_c, z_c \right)$ must satisfy the following constraint:
  \begin{equation}
    \oiint
    {
      \kernelfunction[2]{ \left( x, y, z \right) | \left( x_c, y_c, z_c \right) } \cdot \dif s
    }
    =
    1
    .
    \label{KernelConditionToBeEqualOne}
  \end{equation}

  Equation \eqref{KernelConditionToBeEqualOne} can be rewritten as
  {
    \small
    \begin{multline*}
      \oiint
      {
        \kernelfunction[2]{ \left( x, y, z \right) | \left( x_c, y_c, z_c \right) } \cdot \dif s
      }
      =
      \oiint
      {
        \kernelfunction[2]{ \sphericaldistance \left( \left( x, y, z \right) , \left( x_c, y_c, z_c \right) \right) } \cdot \dif s
      }
      =
      \\
      =
      \sum\limits_{k = 1}^{n}
      {
        \left(
        a_k^2 \cdot \left( \sphericaldiskarea{\left( r_{k} \right)} - \sphericaldiskarea{\left( r_{k - 1} \right)} \right)
        \right)
      }
      =
      1
      .
    \end{multline*}
  }
  
  To satisfy \eqref{KernelConditionToBeEqualOne}, the coefficient $a_j$ in \eqref{KernelFunctionUnNormalized} is replaced with
  \begin{equation*}
    \dfrac
    {
      a_j
    }
    {
      \sqrt
      {
        \sum\limits_{k = 1}^{n}
        {
          \left(
          a_k^2 \cdot \left( \sphericaldiskarea{\left( r_{k} \right)} - \sphericaldiskarea{\left( r_{k - 1} \right)} \right)
          \right)
        }
      }
    }
  \end{equation*}

\begin{figure*} [htb]
  \centering
  \includegraphics[width = 12 cm, keepaspectratio]{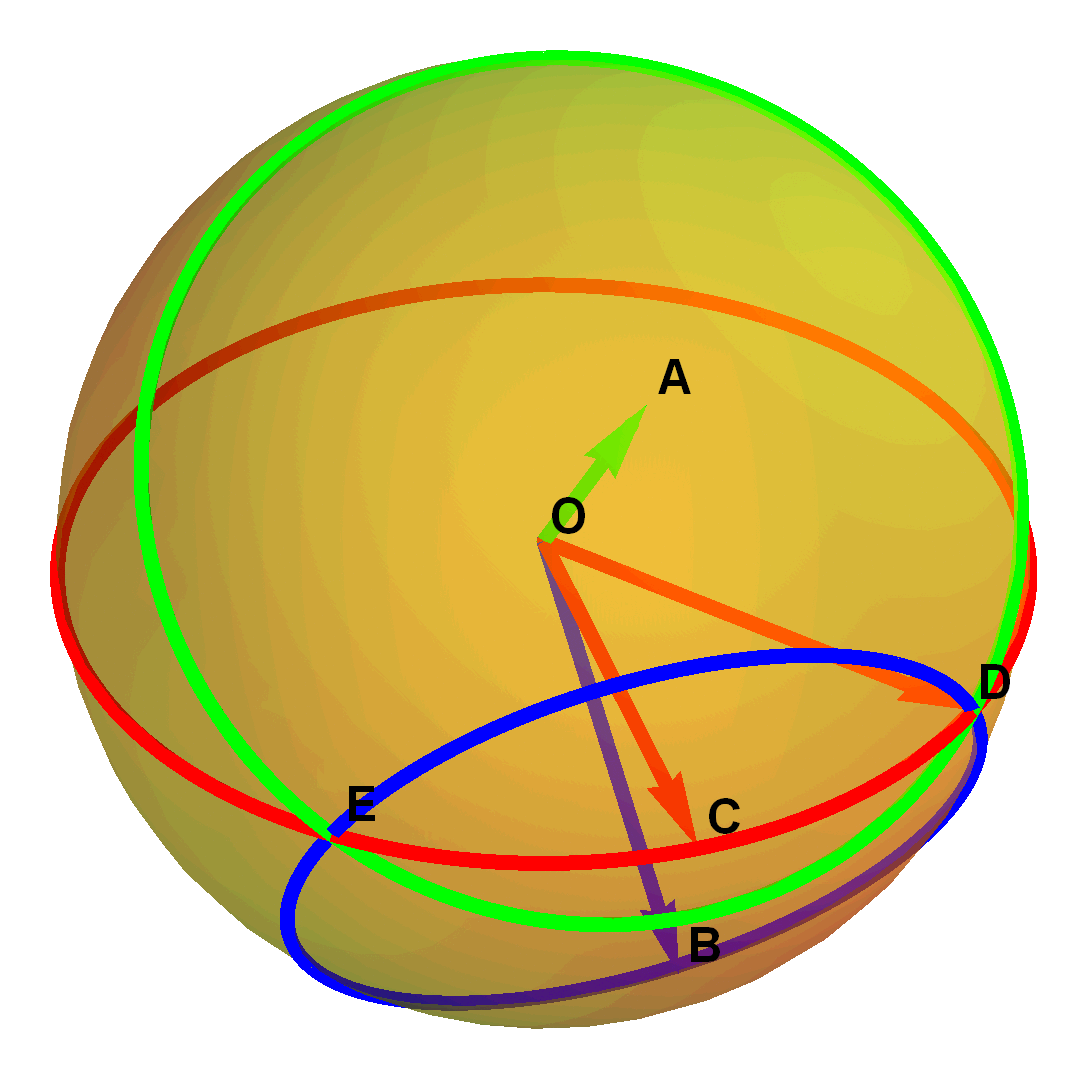}
  \caption{Intersection of the green spherical circle with center $A$ and radius $r_0 = \angle{AOD}$ and the blue spherical circle with center $B$ and radius $r_1 = \angle{BOD}$. Spherical distance between $A$ and $B$ equals $d$. These spherical circles intersect at points $D$ and $E$. $DE$ is a spherical segment. Point $C$ is between $E$ and $D$. $O$ is the center of a sphere.}
  \label{fig:SphereClarification}
\end{figure*}

Next step in the kernel convolution construction is calculation of the intersection area of spherical disks, for example, green and blue spherical disks in Figure~\ref{fig:SphereClarification}.

If a spherical disk is larger than $\dfrac{\pi}{2}$, its complement will be used instead (see \ref{DefinitionComplementOfAnySphericalDisk} and \ref{DefinitionComplementOfAnySphericalDiskMinimumRadius}).

When $r_0 > \dfrac{\pi}{2}$,
\begin{equation}
  \intersectionarea{\left( r_0, r_1, d \right)}
  =
  \sphericaldiskarea{\left( r_1 \right)}
  -
  \intersectionarea{\left(\pi - r_0, r_1, \pi - d \right)}
  .
  \label{IntersectionAreaReverseOnceCircle}
\end{equation}

When $r_1 > \dfrac{\pi}{2}$,
\begin{equation}
  \intersectionarea{\left( r_0, r_1, d \right)}
  =
  \sphericaldiskarea{\left( r_0 \right)}
  -
  \intersectionarea{\left(r_0, \pi - r_1, \pi - d \right)}
  .
  \label{IntersectionAreaReverseOnceCircle2}
\end{equation}

If both spherical radii are larger than $\dfrac{\pi}{2}$, then, from \eqref{IntersectionAreaReverseOnceCircle}, \eqref{IntersectionAreaReverseOnceCircle2}, and \ref{DefinitionSphereArea},
\begin{equation*}
  \intersectionarea{\left( r_0, r_1, d \right)}
  =
  \sphericaldiskarea{\left( r_0 \right)} + \sphericaldiskarea{\left( r_1 \right)} - 4 \pi + \intersectionarea{\left( \pi - r_0, \pi - r_1, d \right)}
  .
\end{equation*}

Therefore, it is sufficient to calculate the intersection area for spherical disks with radii less than or equal to $\dfrac{\pi}{2}$.

If $r_0 + r_1 \leq d$, then the intersection area is equal to zero. If $r_1 \leq r_0 - d$, then the blue spherical disk in Figure~\ref{fig:SphereClarification} is inside the green spherical disk and the intersection area is equal to the area of the blue spherical disk. If $r_0 \leq r_1 - d$, then the green spherical disk is inside the blue spherical disk and the intersection area is equal to the area of the green spherical disk. Otherwise, they intersect each other. In that last case, we need to find the following angles: $\angle{AOC}$, $\angle{BOC}$, $\angle{COD}$, $\sphericalangle{CAD}$, $\sphericalangle{CBD}$, $\sphericalangle{ADC}$, and $\sphericalangle{BDC}$. These angles are calculated as follows:
\begin{equation*}
  \begin{aligned}
    \angle{AOC} = & \left. \dfrac{d}{2} + x \right.,&
    \angle{BOC} = & \left. \dfrac{d}{2} - x \right.,&
  \end{aligned}
\end{equation*}
where
\begin{equation}
  x
  =
  \arctan
  {
    \left(
      \cot{\left( \frac{d}{2} \right)}
      \cdot
      \dfrac
      {
        \sin^2{\left( \frac{r_0}{2} \right)}
        -
        \sin^2{\left( \frac{r_1}{2} \right)}
      }
      {
        1
        -
        \left(
          \sin^2{\left( \frac{r_0}{2} \right)}
          +
          \sin^2{\left( \frac{r_1}{2} \right)}
        \right)
      }
    \right)
  }
  .
  \label{EquationDivisionOfCirclesOnSphere}
\end{equation}

Because $r_0$ and $r_1$ can be close to $\dfrac{\pi}{2}$, care has to be taken when evaluating
$
  \dfrac
  {
    \sin^2{\left( \dfrac{r_0}{2} \right)}
    -
    \sin^2{\left( \dfrac{r_1}{2} \right)}
  }
  {
    1
    -
    \left(
      \sin^2{\left( \dfrac{r_0}{2} \right)}
      +
      \sin^2{\left( \dfrac{r_1}{2} \right)}
    \right)
  }
$.
When both radii equals $\dfrac{\pi}{2}$, any value between $-1$ and $1$ can be used.

This is equivalent to the problem of finding the intersection of disks on the plane. Notice that \eqref{EquationDivisionOfCirclesOnSphere} tends toward
$
  x
  =
  \dfrac
  {
    r_0^2 - r_1^2
  }
  {
    2 d
  }
  ,
$
when  $r_0$, $r_1$, and $d$ approach $0$.

Since $\sphericalangle{ACD}$, $\sphericalangle{ACE}$, $\sphericalangle{BCD}$, and $\sphericalangle{BCE}$ are spherical right angles, it follows from \ref{DefinitionPythagoreanTheorem} that $ \angle{COD} $ can be calculated using one of the following formulas:
%
\begin{equation*}
  \angle{COD}
  =
  2
  \arcsin
  {
    \sqrt
    {
      \dfrac
      {
        \sin^2{\left( \dfrac{r_0}{2} \right)}
        -
        \sin^2{\left( \dfrac{\angle{AOC}}{2} \right)}
      }
      {
        1
        -
        2
        \sin^2{\left( \dfrac{\angle{AOC}}{2} \right)}
      }
    }
  }
  \text{ or }
  \angle{COD}
  =
  2
  \arcsin
  {
    \sqrt
    {
      \dfrac
      {
        \sin^2{\left( \dfrac{r_1}{2} \right)}
        -
        \sin^2{\left( \dfrac{\angle{BOC}}{2} \right)}
      }
      {
        1
        -
        2
        \sin^2{\left( \dfrac{\angle{BOC}}{2} \right)}
      }
    }
  }
  .
\end{equation*}

In the formulas above, the denominator under the square root can be close to zero. Therefore, to avoid numerical instability, the equation with the largest denominator should be used. Note that $\angle{COD}$ is always less than or equal to the minimum of $r_0$ and $r_1$.

Using expression \ref{DefinitionSphericalAngleOfSphericalRightTriangle}, $ \sphericalangle{CAD} $ and $ \sphericalangle{CBD} $ can be calculated as
\begin{equation*}
  \begin{aligned}
    \sphericalangle{CAD} = & \arccos{\left( \cot{\left( r_0 \right)} \cdot \tan{\left( \angle{AOC} \right)} \right)},\\
    \sphericalangle{CBD} = & \arccos{\left( \cot{\left( r_1 \right)} \cdot \tan{\left( \angle{BOC} \right)} \right)}.
  \end{aligned}
\end{equation*}

Next, the intersection area of spherical disks is calculated using \ref{DefinitionAreaOfSphericalCircularSegment} as
\begin{multline}
  \intersectionarea{\left( r_0, r_1, d \right)}
  =
  \\
  =
  \sphericalsegmentarea{\left( \sphericalangle{CAD}, r_0, \angle{COD}, \angle{AOC} \right)}
  +
  \sphericalsegmentarea{\left( \sphericalangle{CBD}, r_1, \angle{COD}, \angle{BOC} \right)}
  .
  \label{FinalFormularIntersectionArea}
\end{multline}

There are two cases when centers of the spherical circles $A$ and $B$ are located on both sides of $ECD$ and when $A$ and $B$ are on the same side. Formula \eqref{FinalFormularIntersectionArea} is correct for both cases. Note that in the case when both centers of the spherical circles $A$ and $B$ are located on the same side of $ECD$, the sign of the area of one of the spherical triangles is negative.

In anisotropic case, to avoid solving a problem of finding intersection of two ellipses on a sphere, the ellipses can be approximated using four spherical arcs as described in \cite{Reference16Oval}.

Finally, the covariance between two locations separated by the spherical distance $d$ is produced by convolution of two kernels. Using \eqref{KernelFunctionUnNormalized} and \eqref{FinalFormularIntersectionArea}, the covariance between locations $ \left( x_1, y_1, z_1 \right) $ and $ \left( x_2, y_2, z_2 \right) $ on a sphere, with spherical distance between points $ \left( x_1, y_1, z_1 \right) $ and $ \left( x_2, y_2, z_2 \right) $ equal to $ d $, is
\begin{equation*}
  \mycovariancefunction{d}
  =
  \oiint
  {
    \kernelfunction{ \left( x, y, z \right) | \left( x_1, y_1, z_1 \right) }
    \cdot
    \kernelfunction{ \left( x, y, z \right) | \left( x_2, y_2, z_2 \right) }
    \cdot
    \dif s
  }
  =
  \sum\limits_{j_1 = 1}^{n}
  {
    \sum\limits_{j_2 = 1}^{n}
    {
      \left(
        b_{j_1}
        \cdot
        b_{j_2}
        \cdot
        \intersectionarea
        {
          \left(
            r_{j_1},
            r_{j_2},
            d
          \right)
        }
      \right)
    }
  }
  =
\end{equation*}
\begin{equation*}
  =
  \sum\limits_{j = 1}^{n}
  {
    \left(
      b_{j}^2
      \cdot
      \intersectionarea
      {
        \left(
          r_{j},
          r_{j},
          d
        \right)
      }
    \right)
  }
  +
  2
  \cdot
  \sum\limits_{j_1 = 1}^{n}
  {
    \sum\limits_{j_2 = 1}^{j_1 - 1}
    {
      \left(
        b_{j_1}
        \cdot
        b_{j_2}
        \cdot
        \intersectionarea
        {
          \left(
            r_{j_1},
            r_{j_2},
            d
          \right)
        }
      \right)
    }
  }
  .
\end{equation*}

\section{Conclusion}

One way to construct a valid covariance model in n\=/dimensional space is by performing ker\-nel\-/con\-vo\-lu\-tion proposed in \cite{Reference1bKernelConvolution} and \cite{Reference1aKernelConvolution}. However, that methodology is not applicable for the data collected on a sphere because there is no regular grid with sufficient number of nodes for applying the fast Fourier transform algorithm as in \cite{KernelConvolutionFFT}. Therefore, we define a kernel function as a set of rings and use these rings instead of rectangles as in \cite{Reference1bKernelConvolution} and \cite{Reference1aKernelConvolution}. For each ring, the kernel function has constant value. Our methodology extends the applicability of the ker\-nel\-/con\-vo\-lu\-tion approach from n\=/dimensional space to the n\=/dimensional sphere.

When the kernel function is represented by a series of rectangles, the resultant covariance function is piece-wise bilinear and anisotropic. On the other hand, when the kernel function is constructed using rings, the resultant covariance function is isotropic, which allows for precalculation of the covariance model for further efficient use for the data fitting, predictions, and simulations.

The algorithm for prediction and simulation with compactly supported covariance on a sphere proposed in this paper can be the following:
\begin{enumerate}
  \item Choose a set of kernels with fixed bandwidth.
  \item Approximate these kernels using a step function.
  \item Approximate the corresponding covariance model by piecewise polynomials and save the tabulated values of the covariance.
  \item For prediction, fit the tabulated covariance, then find the corresponding kernel and perform ker\-nel\-/con\-vo\-lu\-tions kriging or use the covariance models merging as in \cite{Reference9PragmaticBayesianKriging}.
  \item For simulations, approximate kernels at the required locations as in \cite{KernelConvolutionForRingsOnPlane}, generate unconditional simulation using, for example, the approach from \cite{SimulationsOnSphere}, and produce conditional simulation using the relationship between the kernels and covariance functions~\eqref{eq:SphericalCovarianceFunction}.
\end{enumerate}

In practice, spatial data variation is changing from place to place and, in the case of interpolation of large datasets, it is advantageous to divide the data into subset, estimate the covariance model in that subsets and then use one of the available approaches for merging the models, as discussed, for example, in \cite{Reference9PragmaticBayesianKriging} for predictions and in \cite{KernelConvolutionForRingsOnPlane} for simulations.

\section*{Acknowledgement}

The authors would like to thank the anonymous reviewer for valuable comments and suggestions which helped to improve the paper quality.


\newcommand{\doi}[1]{\textsc{doi}: \href{http://dx.doi.org/#1}{\nolinkurl{#1}}}


\bibliographystyle{IEEEtran}

\bibliography{NewFlexibleCompactCovarianceModelOnASphere}

\begin{thebibliography}{10}
\providecommand{\url}[1]{#1}
\csname url@samestyle\endcsname
\providecommand{\newblock}{\relax}
\providecommand{\bibinfo}[2]{#2}
\providecommand{\BIBentrySTDinterwordspacing}{\spaceskip=0pt\relax}
\providecommand{\BIBentryALTinterwordstretchfactor}{4}
\providecommand{\BIBentryALTinterwordspacing}{\spaceskip=\fontdimen2\font plus
\BIBentryALTinterwordstretchfactor\fontdimen3\font minus
  \fontdimen4\font\relax}
\providecommand{\BIBforeignlanguage}[2]{{%
\expandafter\ifx\csname l@#1\endcsname\relax
\typeout{** WARNING: IEEEtran.bst: No hyphenation pattern has been}%
\typeout{** loaded for the language `#1'. Using the pattern for}%
\typeout{** the default language instead.}%
\else
\language=\csname l@#1\endcsname
\fi
#2}}
\providecommand{\BIBdecl}{\relax}
\BIBdecl

\bibitem{Reference5KBesselOnSphere}
\BIBentryALTinterwordspacing
J.~Guinness and M.~Fuentes, ``Covariance functions for mean square
  differentiable processes on spheres,'' August 2013, (preprint). [Online].
  Available:
  \url{http://www.stat.ncsu.edu/information/library/papers/mimeo2652_Guinness.pdf}
\BIBentrySTDinterwordspacing

\bibitem{Reference6AnisotropicProcessOnSphere}
M.~Hitczenko and M.~L. Stein, ``Some theory for anisotropic processes on the
  sphere,'' \emph{Statistical Methodology}, vol.~9, no. 1-2, pp. 211--227,
  2012, special Issue on Astrostatistics + Special Issue on Spatial Statistics.

\bibitem{CompactCovariance}
M.~Liang and D.~Marcotte, ``A class of non-stationary covariance functions with
  compact support,'' \emph{Stochastic Environmental Research and Risk
  Assessment}, vol.~30, no.~3, pp. 973--987, 2015.

\bibitem{KernelConvolutionForRingsOnPlane}
A.~Gribov and K.~Krivoruchko, ``Simulations from spatially varying kriging
  model with compactly supported covariance,'' in \emph{Proceedings of IAMG
  2015, The 17th Annual Conference of the International Association for
  Mathematical Geosciences}, H.~Schaeben, R.~T. Delgado, K.~G. van~den
  Boogaart, and R.~van~den Boogaart, Eds., Freiberg, Germany, September 2015,
  pp. 633--639, (DVD).

\bibitem{Refernce13CovarianceIntegration}
\BIBentryALTinterwordspacing
Y.~Li, ``Non-parametric and semi-parametric estimation of spatial covariance
  function,'' Graduate Theses and Dissertations, p. 100, 2013, paper 13268.
  [Online]. Available: \url{http://lib.dr.iastate.edu/etd/13268}
\BIBentrySTDinterwordspacing

\bibitem{Reference1bKernelConvolution}
\BIBentryALTinterwordspacing
R.~P. Barry and J.~M. Ver~Hoef, ``Blackbox kriging: Spatial prediction without
  specifying variogram models,'' \emph{Journal of Agricultural, Biological, and
  Environmental Statistics}, vol.~1, no.~3, pp. 297--322, September 1996.
  [Online]. Available: \url{http://www.jstor.org/stable/1400521}
\BIBentrySTDinterwordspacing

\bibitem{Reference1aKernelConvolution}
D.~Higdon, ``\BIBforeignlanguage{English}{A process-convolution approach to
  modelling temperatures in the north atlantic ocean},''
  \emph{\BIBforeignlanguage{English}{Environmental and Ecological Statistics}},
  vol.~5, no.~2, pp. 173--190, 1998.

\bibitem{SphericalTrigonometry}
\BIBentryALTinterwordspacing
I.~Todhunter, \emph{Spherical Trigonometry}, 5th~ed.\hskip 1em plus 0.5em minus
  0.4em\relax Project Gutenberg, November 2006. [Online]. Available:
  \url{http://www.gutenberg.org/ebooks/19770}
\BIBentrySTDinterwordspacing

\bibitem{CircularSector}
\BIBentryALTinterwordspacing
E.~W. Weisstein, ``Circular sector. {From MathWorld--A Wolfram Web Resource}.''
  [Online]. Available: \url{http://mathworld.wolfram.com/CircularSector.html}
\BIBentrySTDinterwordspacing

\bibitem{CircularSegment}
\BIBentryALTinterwordspacing
------, ``Circular segment. {From MathWorld--A Wolfram Web Resource}.''
  [Online]. Available: \url{http://mathworld.wolfram.com/CircularSegment.html}
\BIBentrySTDinterwordspacing

\bibitem{Reference16Oval}
P.~L. Rosin, ``A survey and comparison of traditional piecewise circular
  approximations to the ellipse,'' \emph{Computer Aided Geometric Design},
  vol.~16, no.~4, pp. 269--286, 1999.

\bibitem{KernelConvolutionFFT}
\BIBentryALTinterwordspacing
J.~M. Ver~Hoef, N.~Cressie, and R.~P. Barry,
  ``\BIBforeignlanguage{English}{Flexible spatial models for kriging and
  cokriging using moving averages and the fast {F}ourier transform ({FFT})},''
  \emph{\BIBforeignlanguage{English}{Journal of Computational and Graphical
  Statistics}}, vol.~13, no.~2, pp. 265--282, 2004. [Online]. Available:
  \url{http://www.jstor.org/stable/1391176}
\BIBentrySTDinterwordspacing

\bibitem{Reference9PragmaticBayesianKriging}
K.~Krivoruchko and A.~Gribov, ``\BIBforeignlanguage{English}{Pragmatic
  {B}ayesian kriging for non-stationary and moderately non-gaussian data},'' in
  \emph{\BIBforeignlanguage{English}{Mathematics of Planet Earth}}, ser.
  Lecture Notes in Earth System Sciences, E.~Pardo-Ig{\'u}zquiza,
  C.~Guardiola-Albert, J.~Heredia, L.~Moreno-Merino, J.~J. Dur{\'a}n, and J.~A.
  Vargas-Guzm{\'a}n, Eds.\hskip 1em plus 0.5em minus 0.4em\relax Springer
  Berlin Heidelberg, 2014, pp. 61--64.

\bibitem{SimulationsOnSphere}
\BIBentryALTinterwordspacing
L.~V. {Hansen}, T.~L. {Thorarinsdottir}, E.~{Ovcharov}, T.~{Gneiting}, and
  D.~{Richards}, ``{G}aussian random particles with flexible {H}ausdorff
  dimension,'' \emph{ArXiv e-prints}, February 2015. [Online]. Available:
  \url{http://arxiv.org/abs/1502.01750}
\BIBentrySTDinterwordspacing

\end{thebibliography}

\end{document}